\documentclass[12pt]{article}
\usepackage{amsmath}
\usepackage{cite}
\usepackage{relsize}
\usepackage{amssymb}
\usepackage{graphics}
\usepackage{epsfig}
\usepackage{epstopdf}
\usepackage{verbatim}
\usepackage{color}
\usepackage{subcaption}
\usepackage{multirow}
\usepackage{textcomp}
\usepackage{float}
\usepackage[dvipsnames]{xcolor}
\usepackage{mathtools}
\usepackage[toc,page]{appendix}
\usepackage{enumitem}

\begin{document}

\title{\textbf{Dark Matter and $(g-2)_{e,\mu}$ in ISS(2,3) based Gauged $U(1)_{L_{e}-L{\mu}}$ Symmetric Model  }}

\author {\small{Rishu Verma\thanks{ rishuvrm274@gmail.com}, Ankush\thanks{ ankush.bbau@gmail.com}, and B. C. Chauhan\thanks{ bcawake@hpcu.ac.in}}}

\date{\textit{Department of Physics and Astronomical Science,\\Central University of Himachal Pradesh, Dharamshala 176215, INDIA.}}

\maketitle

\begin{abstract}
We proposed a model which can explain the neutrino phenomenology, dark matter and anomalous magnetic moment$(g-2)$ in a common framework. The inverted sea saw (ISS)(2,3) mechanism has been incorporated, in which we get an extra sterile state and this state act as a viable dark matter candidate. The right handed neutrino mass is obtained in TeV scale, which is accessible at LHC. The anomaly free $U(1)_{L_{e}-L{\mu}}$ gauge symmetry is introduced to explain the anomalous magnetic moment of electron and muon because it provides a natural origin of $(g-2)$ in a very minimal setup. The corresponding MeV scale gauge boson successfully explain the anomalous magnetic moment of electron and muon$(g-2)_{e,\mu}$, simultaneously. Thus obtained neutrino phenomenology and relic abundance of dark matter are compatible with experimental results.
\end{abstract}

\section{Introduction}
The Standard Model(SM) of particle physics is an incredible theory which can successfully explain the interactions of fundamental particles and their dynamics. Despite its huge triumph, it lacks in explaining the neutrino mass, matter- antimatter asymmetry, dark matter(DM), anomalous magnetic moment(g-2)$_{e,\mu}$, etc. The experimental discoveries from Super-Kamiokande \cite{SK}, SNO \cite{SNO,SNO1} and KamLand \cite{KAM} has confirmed the solar and atmospheric neutrino flavour oscillations and massive nature of neutrinos. There are many other open questions in neutrino physics like, absolute mass scale of neutrinos, their mass hierarchy(Normal Ordering or Inverted Ordering), whether they are Majorana or Dirac, and CP violation etc., which are required to be answered. Certainly, all of these issues require a framework beyond SM(BSM).

\noindent There are several mechanisms in literature to generate the mass of neutrinos. The simplest and most popular way is to add right handed neutrino (RHN) to the existing SM by hand. In this way the Higgs can have the required Yukawa coupling with the neutrinos. To achieve this, there are several seesaw mechanisms like, Type-I \cite{Seesaw1, Seesaw11}, Type-II \cite{Seesaw21}, Type-III \cite{Seesaw3, Seesaw31} and Inverse Seesaw(ISS) \cite{ISS1,ISS2,ISS3}.
 
 \noindent Another important unsolved problem in cosmology is that of dark matter and its nature. Various astronomical and cosmological experiments suggest that the universe consists of non-luminous, non-baryonic, mysterious matter called dark matter \cite{DM, DM1}. There are several evidences which confirms the presence of such unseen tangible structures in the universe. The cosmic microwave background(CMB), gravitational lensing and galactic rotation curves are some of these evidences \cite{Evidence1,Evidence2,Evidence3}. Despite of such strong evidences, the nature of DM and its origin is still an open question. The current abundance of DM according to Plank is reported as \cite{PLANCK}
 \begin{center}
$\Omega_{DM}h^2 = 0.120\pm0.001$.
\end{center}
 Any particle to be a DM candidate should fulfill the criteria given in \cite{DM2}. According to these specifications, all the existing particles of SM are clearly ruled out to be a DM candidate. Hence in order to obtain correct DM phenomenology, we need a physics beyond the SM.
 
 \noindent Further, the results from LSND \cite{LSND}, MINOS \cite{MINOS} etc, favour the existence of fourth state of neutrino known as, sterile neutrino. The anomalies found in the experiments like GALLEX \cite{GALLEX} and SAGE \cite{SAGE} strengthen this fact, as they are successfully explained by incorporating the sterile neutrino. Sterile neutrino is a right handed neutrino and and can sense only the gravitational interaction. These neutrinos can be produced by their mixing with the active neutrino sector \cite{Ster1,Ster2}. Sterile neutrino is also, one of the popular warm dark matter(WDM) candidate having very small mixing with SM neutrinos leading to its stability \cite{Ster3}. It is a massive particle having very long lifetime. Sterile neutrinos having masses in $keV$ range are extremely important in explaining the cosmological findings \cite{Ster4}. According to cosmological predictions, the DM candidate should have mass range $0.4keV<<m_{DM}<<50keV$. The lowest bound is obtained in \cite{Ster5} and the upper bound is given in \cite{Ster6,Ster7}. Above 50$keV$ mass, the sterile neutrino loses its stability.
 
 \noindent The magnetic moment of charged leptons plays an important role as it is a viable test for the theory of SM. The results from Fermi National Accelerator Laboratory (FNAL) has confirmed that the experimental value of magnetic moment of muon is not compatible with the SM model prediction with 4.2$\sigma$ discrepancy \cite{FNAL}. Similarly, the experimental value of magnetic moment of electron has 1$\sigma$ and 2.4$\sigma$ discrepancy over SM from Rubidium atom and Cesium atom measurements, respectively \cite{rubidium,cesium}. These results also open a new window of BSM physics.
 
\noindent Motivated by all these studies explained above, we explored the possibility of a common phenomena, which can explain the neutrino and DM phenomenology as well as anomalous magnetic moment of electron and muon $(g-2)_{e,\mu}$. We develop a model, in which we used ISS(2,3) framework to generate non-zero neutrino masses. The motivation for incorporating this framework is the minimal formalism, which can generate the right handed neutrino masses in TeV scale, which are accessible at LHC and provide a viable DM candidate, simultaneously \cite{nayana}. We used anomaly free $U(1)_{L_{e}-L{\mu}}$ gauge symmetry, so that we can explain the anomalous magnetic moment of electron and muon, which provides a natural origin of (g-2) in a very minimal setup. Here, we also implemented Type-II seesaw to explain all these phenomena simultaneously. The SM is extended by two RHN($N_1, N_2$) and three singlet sterile neutrinos $S_i$(i = 1,2,3) as required in ISS(2,3) framework. The field content of SM is further extended by three extra fields $\phi$(scalar singlet), $\eta$(scalar doublet) and $\Delta$(scalar triplet). Only $\phi$ will have its contribution towards $(g-2)_{e,\mu}$. Finally, the cyclic symmetry $Z_3$ has also been added to have economical formulation of mass matrices. 
 
\noindent This paper is organised as follow: In section \ref{sec:2} we have given a detailed explanation of the ISS(2,3) formalism. In section \ref{sec:3}, the model part is explained and the effective neutrino mass matrix is obtained. Section \ref{sec:4} contains the discussion of DM in ISS(2,3) mechanism. Anomalous magnetic moment of muon and electron is explained in Section \ref{sec:5}. Numerical analysis and results are presented in Section \ref{sec:6}. Finaly, the conclusions are given in Section \ref{sec:7}.

\section{Inverse Seesaw(2,3) Formalism}\label{sec:2}
\noindent In order to explain the smallness of neutrino masses, there are different seesaw mechanisms explained in literature \cite{Seesaw1,Seesaw21,Seesaw3}. In canonical seesaw models, we obtain the right handed neutrino mass around GeV scale. But, ISS provides a formalism in which one can obtain the right handed neutrino mass at TeV scale, which can be probed at LHC and other future experiments. The ISS(2,3) is the minimal formalism to obtain the dark matter(DM) candidate \cite{ISS231, ISS232} as well as neutrino phenomenology. This is possible due to the fact that in this scenario we have unequal number of right-handed(RH) neutrinos $N_j(j =1,2)$ and singlet fermions $S_i(i =1,2,3)$, which leads to a DM candidate and two pseudo-Dirac pairs \cite{ISS233}. The mass Lagrangian in conventional ISS mechanism is written as
\begin{equation}
L = -\bar{\nu}_{\alpha L}M_{D}N_{j} - \bar{S_{i}}MN_{j}-\frac{1}{2}\bar{S_{i}}\mu S^C_{k} + h.c., \label{eq:1}
\end{equation}
where $M_D$, $M$ and $\mu$ are the complex mass matrices and $\alpha = (e,\mu,\tau)$, $k=(1,2,3)$. After spontaneous symmetry breaking, the above Eq.(\ref{eq:1}) becomes following $9\times9$ neutrino mass matrix
\begin{equation}
 M_{\nu} = \begin{pmatrix}
0 & M_{D} & 0\\
M_D^T & 0 & M\\
0 & M^T & \mu \\
\end{pmatrix}, \\
\label{eq:2}
\end{equation}

\noindent where $M_D$ is the dirac mass matrix, $M$ represents the interaction of the RH neutrinos with singlet fermions and $\mu$ represents the mass matrix of singlet fermions. The SM neutrinos are obtained at sub-eV scale from $M_D$,  $\mu$ at keV scale and M at TeV scale \cite{ISS1,ISS2,ISS3}. Considering $\mu << M_D << M$, after block diagonalization of the matrix in Eq.(\ref{eq:2}), the 3$\times$3 effective neutrino matrix is given as

\begin{equation}
 m_{\nu} \approx M_{D} (M{^T})^{-1}\mu M^{-1} M_{D}^{T}.\\
 \label{eq:3}
\end{equation}

and the mass matrix for heavy sector can be written as \cite{ISS234}
\begin{equation}
 M_{H} = \begin{pmatrix}
0 & M \\
M^T & \mu\\
\end{pmatrix}, \\
\label{eq:4}
\end{equation}

\noindent where $M_{H}$ represents the mass matrix for heavy pseudo-Dirac pairs and extra fermions. Since, in ISS(2,3) scenario, the $M$ is a 2$\times$3 matrix. Therefore, it is not possible to calculate $M^{-1}$. Consequently, to obtain effective neutrino mass matrix we used the formalism as given in Ref.\cite{ISS235}
\begin{equation}
 m_{\nu} \approx M_{D} d  M_{D}^{T},\\
 \label{eq:5}
\end{equation}
where d is $2\times2$ matrix,
\begin{equation}
 M^{-1}_{H} = \begin{pmatrix}
d_{2\times2} & ... \\
... & ...\\
\end{pmatrix}. \\
\label{eq:6}
\end{equation}

  \section{The Model}\label{sec:3}

\noindent In the model, we have included two right-handed neutrinos $N_j$(j=1,2) and three singlet fermions $S_i(i = 1,2,3)$, which are charged ($-1, 1$) and $(0,0,0)$ under $U(1)_{L_{e}-L{\mu}}$ , respectively. Further, we extended the scalar sector with one $SU(2)_L$ singlet scalar field $\phi$ and a scalar doublet $\eta$ which are charged $1$ and $-1$ respectively under $U(1)_{L_{e}-L{\mu}}$.  

\noindent The scalar field $\phi$ is breaking the $U(1)_{L_{e}-L{\mu}}$ symmetry, while H(Higgs doublet) is responsible for breaking electroweak symmetry. After spontaneous symmetry breaking (SSB), the vacuum expectation values (VEV) acquired by ($H$ and $\eta$) and ($\phi$) give $M_D$ and $M$, respectively, with minimal number of parameters.  In addition, $Z_3$ symmetry is used to constrain the Yukawa Lagrangian. The fermionic and scalar field content along with respective charge assignments are shown in Table \ref{table1}.

\begin{center}
\begin{table}[h]
\centering
\begin{tabular}{cccccccccccccc}
 Symmetry & $\bar{L}_{e}$ & $\bar{L}_{\mu}$ & $\bar{L}_{\tau}$ & $e_{R}$ & $\mu_{R}$ & $\tau_{R}$ & $N_{1}$ & $N_{2}$ & $S_i$ & $H$ & $\phi$ & $\eta$ & $\Delta$ \\
 \hline
$SU(2)_L$    &     2  & 2 & 2 & 1  & 1 & 1   & 1   &   1     &    1        &   2        &    1 &2  & 3 \\

\hline
$U(1)_{L_{e}-L{\mu}}$ & 1  & -1 & 0  &   -1    & 1  &  0  & -1 & 1 &  0 & 0 & 1 & -1 & -1 \\
\hline
$Z_{3}$ & $\omega$  & $\omega$ & $\omega$  &   $\omega^2$    &$\omega^2$  & $\omega^2$  & $\omega^2$ & $\omega^2$ &  1 & 1 & $\omega$ & 1 & $\omega^2$ \\

\hline
\end{tabular}
 \caption{The fermionic and scalar field content with respective charge assignments under $SU(2)_L$, $U(1)_{L_{e}-L{\mu}}$ and $Z_{3}$ symmetries.}
 \label{table1}
\end{table}
\end{center}

\noindent The leading Yukawa Lagrangian is
\begin{eqnarray}
\nonumber
\mathcal{L}^I =&&y_{e}\Bar{L}_{e}e_{R}H + y_{\mu}\Bar{L}_{\mu}\mu_{R}H + y_{\tau}\Bar{L}{_\tau}\tau_{R}H + y_{1}^{\nu}\Bar{L}_{e}N_{1} \Tilde{H} + y_{2}^{\nu}\Bar{L}_{\mu}N_{2} \Tilde{H} +\\
&&  y_{3}^{\nu}\Bar{L}_{\tau}N_{1} \Tilde{\eta}+y_{\phi}^{1} N_{1} S_{1} \phi+ y_{\phi}^{2} N_{1} S_{2} \phi + y_{\phi}^{3} N_{1}S_{3} \phi+p_i S_{i}S_{i} + h.c.,
 \label{eq:7}
\end{eqnarray}

\noindent where $\Tilde{H}=i\tau_{3}H$ and $y_q(q = e,\mu,\tau)$, $y_{i}^{\nu} (i = 1,2,3)$, $y_{\phi}^{j}(j=1,2,3)$ are Yukawa coupling constants. 
The VEVs 
\begin{center}
$\langle H \rangle = v_{H}$,  
$\langle \phi \rangle = v_{\phi}$ and $\langle \eta \rangle = v_{\eta}$.
\end{center}

 \noindent lead to diagonal charged lepton mass matrix as\begin{equation}
m_{l}= Diag(y_{e},y_{\mu},y_{\tau})v_{H}.\\
\end{equation}   
 
\noindent Consequently, we have obtained $M_{D}$, $M$ and $\mu$ as shown below \\
\begin{equation}
M_{D} =\begin{pmatrix}
a & 0 \\
0 & b \\
m & 0 \\
 \end{pmatrix}, M = \begin{pmatrix}
A & B & G \\
B & 0 & 0  \\

\end{pmatrix} ,
\mu = \begin{pmatrix}
p_1 & 0 & 0\\
0 & p_2 & 0 \\
0 & 0 & p_3 \\
\end{pmatrix}, \\
\end{equation}

\noindent where $a=y_{1}^{\nu}v_{H}, b = y_{3}^{\nu}v_{H},m = y_{3}^{\nu}v_{\eta}, A = y_{\phi}^{1}v_{\phi}, B =  y_{\phi}^{2}v_{\phi},G =  y_{\phi}^{3}v_{\phi}$. For numerical estimation, we have assumed the degenerate masses for sterile singlet fermions, which results in lowest mass state eigenstate of heavy sector to be of keV range($p_1\approx p_2\approx p_3 \approx p$). Within ISS(2,3) mechanism, the above matrices lead to the light neutrino mass matrix as follow\\
\begin{equation}
 m_{\nu}^1 = \begin{pmatrix}
\frac{-a^2p}{(B^2+G^2)} & \frac{aAbp}{(B^3+BG^2)} & \frac{-amp}{(B^2+G^2)} \\
\frac{aAbp}{(B^3+BG^2)} &\frac{-b^2(A^2+B^2+G^2)p}{B^2(B^2+G^2)} & \frac{Abmp}{(B^3+BG^2)} \\
\frac{-amp}{(B^2+G^2)} & \frac{Abmp}{(B^3+BG^2)} & \frac{-m^2p}{(B^2+G^2)} \\
\end{pmatrix}.
\label{eq:8}\\
\end{equation}
 \noindent We implemented type-II seesaw to get correct neutrino  phenomenology. To implement type-II seesaw, we introduced a $SU(2)_L$ triplet field $\Delta$ in the model transforming as (-$1$) and $\omega^2$ under $U(1)_{L_{e}-L{\mu}}$ and  $Z_3$, respectively. The relevant Lagrangian corresponding to type-II seesaw is given by
\begin{eqnarray}
 \mathcal{L^{II}} = &&f_{1}(L_{\mu}L_{\tau}+L_{\tau}{L}_{\mu})\Delta + h.c.,
 \label{eq:9}
 \end{eqnarray}
 \noindent where, $f_1$ is coupling constant. The vacuum expectation value $ \langle \Delta \rangle = v_{\Delta} $ gives
\begin{equation}
m_{{\nu}_{II}} = \begin{pmatrix}
0 & 0 & 0 \\
0 & 0 & X \\
0 & X & 0 \end{pmatrix},
\end{equation}

\noindent where, $X = f_1 v_{\Delta}$. The complete Lagrangian for the model is given as

\begin{eqnarray}
\nonumber
\mathcal{L} =&&y_{e}\Bar{L}_{e}e_{R}H + y_{\mu}\Bar{L}_{\mu}\mu_{R}H + y_{\tau}\Bar{L}{_\tau}\tau_{R}H + y_{1}^{\nu}\Bar{L}_{e}N_{1} \Tilde{H} + y_{2}^{\nu}\Bar{L}_{\mu}N_{2} \Tilde{H} +\\
 &&  y_{\phi}^{1} N_{1} S_{1} \phi+ y_{\phi}^{2} N_{1} S_{2} \phi + y_{\phi}^{3} N_{1}S_{3} \phi+p_i S_{i}S_{i}  + f_{1}(L_{\mu}L_{\tau}+L_{\tau}{L}_{\mu})\Delta+ h.c..
 \label{eq:11}
\end{eqnarray}

 The effective neutrino mass matrix is given by
 \begin{equation}
  \nonumber
  M_{\nu} = m_{{\nu}_{I}}+m_{{\nu}_{II}},
\end{equation}
 which explicitly can be written as
\begin{equation}
 M_{\nu} = \begin{pmatrix}
\frac{-a^2p}{(B^2+G^2)} & \frac{aAbp}{(B^3+BG^2)} & \frac{-amp}{(B^2+G^2)} \\
\frac{aAbp}{(B^3+BG^2)} &\frac{-b^2(A^2+B^2+G^2)p}{B^2(B^2+G^2)} & \frac{Abmp}{(B^3+BG^2)}+X \\
\frac{-amp}{(B^2+G^2)} & \frac{Abmp}{(B^3+BG^2)}+X & \frac{-m^2p}{(B^2+G^2)} 
\end{pmatrix} .
\label{eq:12}
\end{equation}

\section{Dark Matter in ISS(2,3) mechanism}\label{sec:4}
As stated earlier, the ISS(2,3) is a mechanism, which can explain  neutrino phenomenology as well as DM, simultaneously. In the generic ISS realization, one can have the following mass spectrum \cite{ISS231} :\\ 
1) scale corresponding to heavy neutrino states of  $\mathcal{O}$($M_D$) and $\mathcal{O}$($M$).\\
2) $\mathcal{O}$($\mu$) scale corresponding to light sterile state ($\#s-\#\nu_R$). This state exists if $\#s < \#\nu_R$ (\# represents the number of neutrinos).  \\
3)Three light active neutrino states of mass scale  $\mathcal{O}$($\mu$) $\frac{k}{1+k^2}$, $k = \frac{\mathcal{O}(M_D)}{\mathcal{O}(M)} $. \\
Using Eq.(\ref{eq:4}) , the $M_H$ in our model is:
\begin{equation}
M_{H} = \begin{pmatrix}
0 & 0 & A & B & G \\
0 & 0 & B & 0 & 0 \\
A & B & p & 0 & 0 \\
B & 0 & 0 & p & 0\\
G & 0 & 0 & 0 & p\end{pmatrix}.
\end{equation}

\noindent The eigen values of $M_H$ are obtained as
\begin{equation}
\nonumber
(M_H)_1 = \frac{1}{4}(2p-2\sqrt{ 2A^2+4B^2+2G^2-2\sqrt {A^4+4A^2B^2+2A^2G^2+G^4}+p^2})    
\end{equation}
\begin{equation}
\nonumber
(M_H)_2 = \frac{1}{4}(2p+2\sqrt{ 2A^2+4B^2+2G^2-2\sqrt {A^4+4A^2B^2+2A^2G^2+G^4}+p^2})    
\end{equation}
\begin{equation}
\nonumber
(M_H)_3 = \frac{1}{4}(2p-2\sqrt{ 2A^2+4B^2+2G^2+2\sqrt {A^4+4A^2B^2+2A^2G^2+G^4}+p^2})    
\end{equation}
\begin{equation}
\nonumber
(M_H)_4 = \frac{1}{4}(2p+2\sqrt{ 2A^2+4B^2+2G^2+2\sqrt {A^4+4A^2B^2+2A^2G^2+G^4}+p^2})    
\end{equation}
\begin{equation}
(M_H)_5 = p    
\end{equation}

\noindent It is to be noted that, $(M_H)_5$ depends only on the $\mu$ matrix. Hence $(M_H)_5$ is the lightest sterile state acting as potential DM candidate. So, the mass of the DM particle can be determined by the `$p$' parameter of the model. In order to study the DM phenomenology, it is important to calculate the active-sterile mixing. This can be obtained from the first three eigenvectors corresponding to the eigenvalues $M$ in keV range. The relation between DM mass and active sterile mixing with relic abundance is given as \cite{relic}\\
\begin{equation}
\Omega_{DM}h^2 = 1.1\times 10^7\Sigma C_{\alpha}(m_s)\mathcal|{U}_{\alpha s}|^2 \left(\frac{m_s^2}{keV}\right),\alpha = e,\mu,\tau
\end{equation}

\noindent The simplified solution of above equation is
\begin{eqnarray}
\Omega_{DM}h^2 = 0.3\left(\frac{sin^2\theta}{10^{-10}}\right) \left(\frac{m_s}{100keV}\right)^2,
\end{eqnarray}
where, $sin^2 2\theta = 4 \Sigma_{\alpha e, \mu, \tau}|{U}_{\alpha s}|^2$. ${U}_{\alpha s}$ represents the active-sterile mixing element and $m_{s}$ is the mass of lightest sterile fermion. The DM particle should be stable atleast at the cosmological scale. The lightest sterile neutrino may decay into active neutrinos and photon $\gamma$,
which leads to the monochromatic X-ray line signal. Its decay rate is negligible as compared to the cosmological scalez due to very small mixing angle and is given as \cite{decay}
\begin{eqnarray}
\Gamma = 1.38 \times 10^{-32}\left(\frac{sin^22 \theta}{10^{-10}}\right)\left(\frac{ms}{keV}\right)^5s^{-1}.
\end{eqnarray}
\noindent The model parameters obtained in the model are used to find the relic and decay rate.

\section{Anomalous Magnetic Moment of Electron and Muon}\label{sec:5}
\begin{figure}[ht]
	\begin{center}
			{\epsfig{file=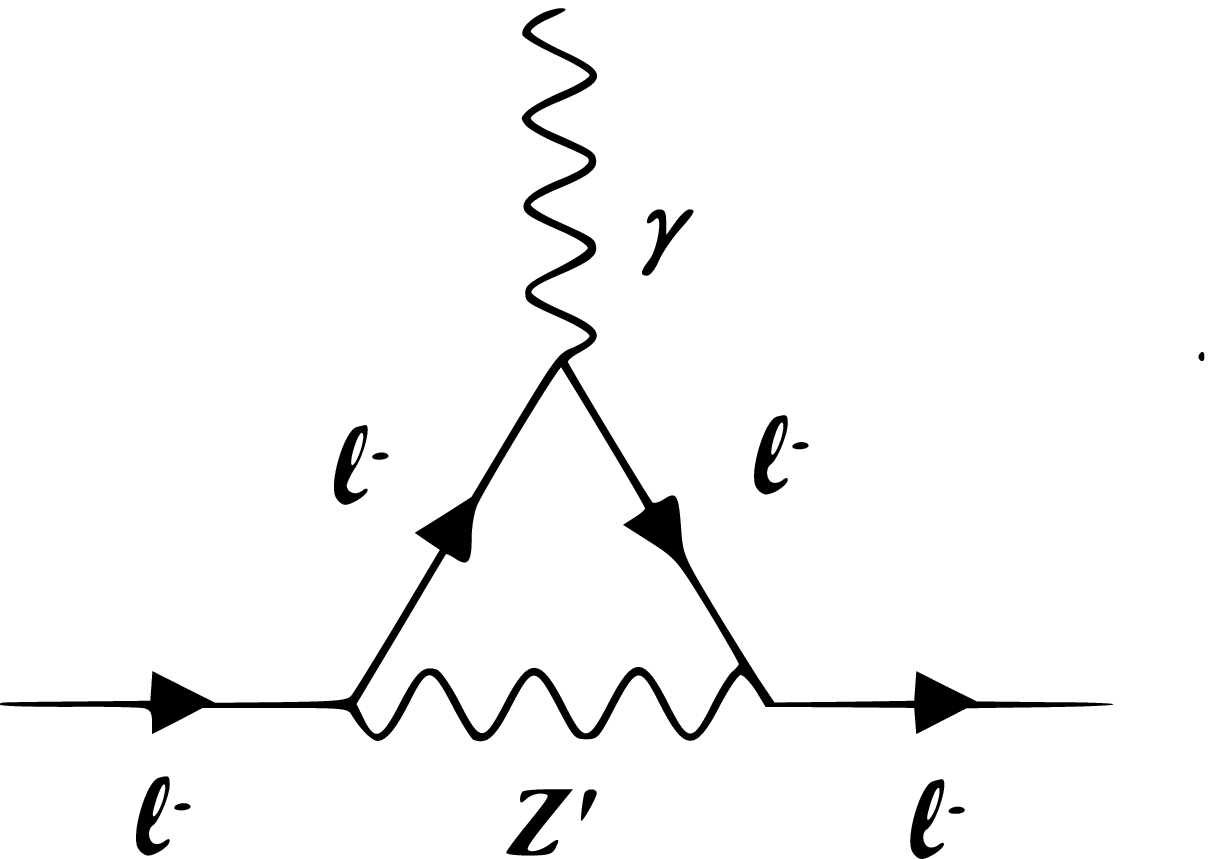,height=4.5cm,width=7.0cm}}
		\end{center}
\caption{\label{fig2}Feynman diagram for electron and muon  (g-2)($l=e,\mu$) with mediator Z$'$ gauge boson.}
\label{Fig2}
\end{figure}

\subsection{Muon (g-2) Anomaly}
The magnetic moment $\Vec{\mu}$ for any elementary particle having charge `$q$' and spin $\Vec{S}$ is given as
\begin{equation}
    \Vec{\mu} = g\frac{q}{2m}\Vec{S},
\end{equation}
\noindent where $m$ and $g$ are the mass of the particle and gyromagnetic ratio, respectively. Using quantum mechanics formulation, Dirac predicted the value of $g$ for spin-1/2 particle to be equal to 2 \cite{Dirac}. The higher order radiative corrections tends to deviate the value of $g$ from 2. In particular, the fractional deviation of $g$ from Dirac's prediction is known as anomalous magnetic moment($a$). For muon the anomalous magnetic moment is defined as $a_{\mu} = (g-2)/2.$ The recent results of muon (g-2) experiment from Fermi National Accelerator Laboratory(FNAL) \cite{FNAL} predicts the value of $a_{\mu}$ as
\begin{equation}
    a_{\mu}^{FNAL} = 116592040(54)\times10^{-11}.
    \label{eq:21}
\end{equation}
On the other hand, the theoretical predictions of SM results in \cite{SM}
\begin{equation}
    a_{\mu}^{SM} = 116591810(43)\times10^{-11}.
    \label{eq:22}
\end{equation}
The 4.2$\sigma$ significance discrepancy $ \Delta a_{\mu} ( a_{\mu}^{FNAL}-a_{\mu}^{SM}$), challenges the SM and hints towards the new physics beyond SM. From Eq.(\ref{eq:21}) and Eq.(\ref{eq:22}), we get
\begin{equation}
    \Delta a_{\mu} = 251(59)\times10^{-11}.
\end{equation}
\noindent There are various models which have discussed $(g-2)_{\mu}$ anomaly can be found in the literature \cite{g1,g2,g3,g4,g5,g6,g7,g8,g9,g10,g11}. 
Since we have extended our model by $U(1)_{L_{e}-L{\mu}}$ symmetry, the $Z'$ boson can contribute to $(g-2)_\mu$ anomaly if its mass is in the range (10-100) MeV. Through the interaction of muon with $Z'$ boson can provide substantial rectification to $g-2$.  The neutral current interaction which gives the contribution to the calculation of $(g-2)_{\mu}$ is given as
\begin{equation}
    \mathcal{L} = - g'\bar\mu Z' \gamma^{\mu} \mu,
\end{equation}
where $g'$ is corresponding coupling constant. The analytical one loop contribution of $Z'$ can be written as\cite{analytical1,analytical2}
\begin{equation}
    \Delta a_{\mu}^{Z'} = \frac{g'^2}{8\pi^2}\int_{0}^{1}dx\frac{2m_{\mu}^2 x^2(1-x)}{x^2m_{\mu}^2+(1-x)M_{Z'}^2},
\end{equation}

where $m_{\mu}$ is the mass of muon and $M_{Z'}$ is the gauge boson mass.
\subsection{Electron (g-2) Anomaly}
Unlike the magnetic moment of muon, the recent experiments are not able to accurately measure the magnetic moment of electron. According to Rubidium atom measurement \cite{rubidium}
\begin{equation}
    (\Delta a_{e})_{Rb} = 48(30)\times10^{-14},
\end{equation}

\noindent with $1\sigma$ discrepancy over SM, whereas, Cesium atom gives
\begin{equation}
    (\Delta a_{e})_{Cs} = -87(30)\times10^{-14},
\end{equation}
\noindent with $2.4\sigma$ discrepancy over SM \cite{cesium}. 
 The neutral current interaction which gives the contribution to the calculation of $(g-2)_e$ is given as
\begin{equation}
    \mathcal{L} = - g'\bar e Z' \gamma^{e} e.
\end{equation}
  The one loop contribution of $Z'$ to the magnetic moment of electron is given as \cite{electron}
\begin{equation}
    \Delta a_{e}^{Z'} = \frac{g'^2}{8\pi^2}\int_{0}^{1}dx\frac{2m_{e}^2 x^2(1-x)}{x^2m_{e}^2+(1-x)M_{Z'}^2},
\end{equation}

\noindent where $m_{e}$ is mass of electron. The contributing Feynman diagram is shown in Fig.\ref{Fig2}. In order to explain these discrepancies in electron and muon anomalous magnetic moment, anomaly free gauge symmetry $U(1)_{L_{e}-L{\mu}}$ is used. The extra gauge boson $Z'$ (in MeV range) obtained from $U(1)_{L_{e}-L{\mu}}$ symmetry breaking effectively contributes to $\Delta a_{\mu}$ and $\Delta a_{e}$ .

\section{Numerical Analysis and Results}\label{sec:6}
In this section we have numerically estimated the viability of the model with the neutrino oscillation data, DM relic and bounds on $\Delta a_{e,\mu}$. From the light neutrino mass matrix $M_{\nu}$ acquired by using ISS(2,3) and type-II seesaw mechanisms in Eq.(\ref{eq:12}), it is clear that we have eight unknown model parameters. For numerical analysis, these model parameters are evaluated using the constraints of neutrino oscillation data as shown in Table \ref{table4}.

\begin{table}[ht]
\begin{center}
\begin{tabular}{c|c|c}
\hline \hline 
Parameter & Best fit $\pm$ \( 1 \sigma \) range & \( 3 \sigma \) range  \\
\hline \multicolumn{2}{c} { Normal neutrino mass ordering \( \left(m_{1}<m_{2}<m_{3}\right) \)} \\
\hline \( \sin ^{2} \theta_{12} \) & $0.304^{+0.013}_{-0.012}$ & \( 0.269-0.343 \)  \\
\( \sin ^{2} \theta_{13} \) & $0.02221^{+0.00068}_{-0.00062}$ & \( 0.02034-0.02420 \) \\
\( \sin ^{2} \theta_{23} \) & $0.570^{+0.018}_{-0.024}$ & \( 0.407-0.618 \)  \\
\( \Delta m_{21}^{2}\left[10^{-5} \mathrm{eV}^{2}\right] \) & $7.42^{+0.21}_{-0.20}$& \( 6.82-8.04 \) \\
\( \Delta m_{31}^{2}\left[10^{-3} \mathrm{eV}^{2}\right] \) & $+2.541^{+0.028}_{-0.027}$ & \( +2.431-+2.598 \) \\
\hline \multicolumn{2}{c} { Inverted neutrino mass ordering \( \left(m_{3}<m_{1}<m_{2}\right) \)} \\
\hline \( \sin ^{2} \theta_{12} \) & $0.304^{+0.013}_{-0.012}$ & \( 0.269-0.343 \)\\
\( \sin ^{2} \theta_{13} \) & $0.02240^{+0.00062}_{-0.00062}$ & \( 0.02053-0.02436 \) \\
\( \sin ^{2} \theta_{23} \) & $0.575^{+0.017}_{-0.021}$& \( 0.411-0.621 \) \\
\( \Delta m_{21}^{2}\left[10^{-5} \mathrm{eV}^{2}\right] \) & $7.42^{+0.21}_{-0.20}$ & \( 6.82-8.04 \) \\
\( \Delta m_{32}^{2}\left[10^{-3} \mathrm{eV}^{2}\right] \) & $-2.497^{+0.028}_{-0.028}$ & \( -2.583--2.412 \)  \\
\hline \hline
\end{tabular}
\end{center}
\caption{Neutrino oscillations experimental data NuFIT 5.0 used in the numerical analysis\cite{data}.}
\label{table4}
\end{table}

\noindent In charged lepton basis, the light neutrino mass matrix can be written as\\
\begin{equation}
M_{\nu}=UM_{d}U^T,
\end{equation}
where $M_{d}$ is diagonal mass matrix containing mass eigenvalues of neutrinos\\
$diag(m_{1}, m_{2}, m_{3})$. 
$ U$ is Pontecorvo-Maki-Nakagawa-Sakata $(U_{PMNS})$ neutrino mixing matrix defined as $U_{PMNS}=V.P$ 
where $ P$ is diagonal phase matrix  $diag(1,e^{i\alpha},e^{i(\beta+\delta)})$, in which $\alpha$, $\beta$ are Majorana type $CP$ violating phases. In PDG representation, $V$ is given by
  \begin{equation}
   \begin{pmatrix}
c_{12} c_{13} & s_{12} c_{13} &  s_{13} e^{-i\delta} \\
-s_{12} c_{23} - c_{12} s_{23} s_{13} e^{i\delta} & c_{12} c_{23} - s_{12} s_{23} s_{13} e^{i\delta} &  s_{23} c_{13} \\
s_{12} s_{23} - c_{12} c_{23} s_{13} e^{i\delta} & -c_{12} s_{23} -s_{12} c_{23} s_{13} e^{i\delta} &  c_{23} c_{13} \\
 \end{pmatrix},
   \end{equation}
\noindent where $\delta$ is Dirac $CP$ violating phase. The oscillation parameters are given as\\

$sin^2 \theta_{13} = |U_{e3}|^2$, \hspace{1cm} $sin^2 \theta_{23} = \frac{|U_{\mu 3}|^2}{1-|U_{e 3}|^2}$\hspace{1cm} and \hspace{1cm} $sin^2 \theta_{12} = \frac{|U_{e2}|^2}{1-|U_{e 3}|^2}$\\

\noindent where $U_{\alpha i}$($\alpha$ = e,$\mu$,$\tau$, and i = 1,2,3) are the elements of $U_{PMNS}$ matrix. To check the viability of the model, we used the neutrino oscillation data as given in Table \ref{table4}. Further, we found the parameter space of the model satisfying the neutrino oscillation data. The model parameter space compatible with the neutrino oscillation data is used to study active-sterile mixing, decay rate of DM candidate and relic abundance of DM.

\begin{figure}[ht]
 \begin{subfigure}[b]{0.4\textwidth}
 \includegraphics[scale=.7]{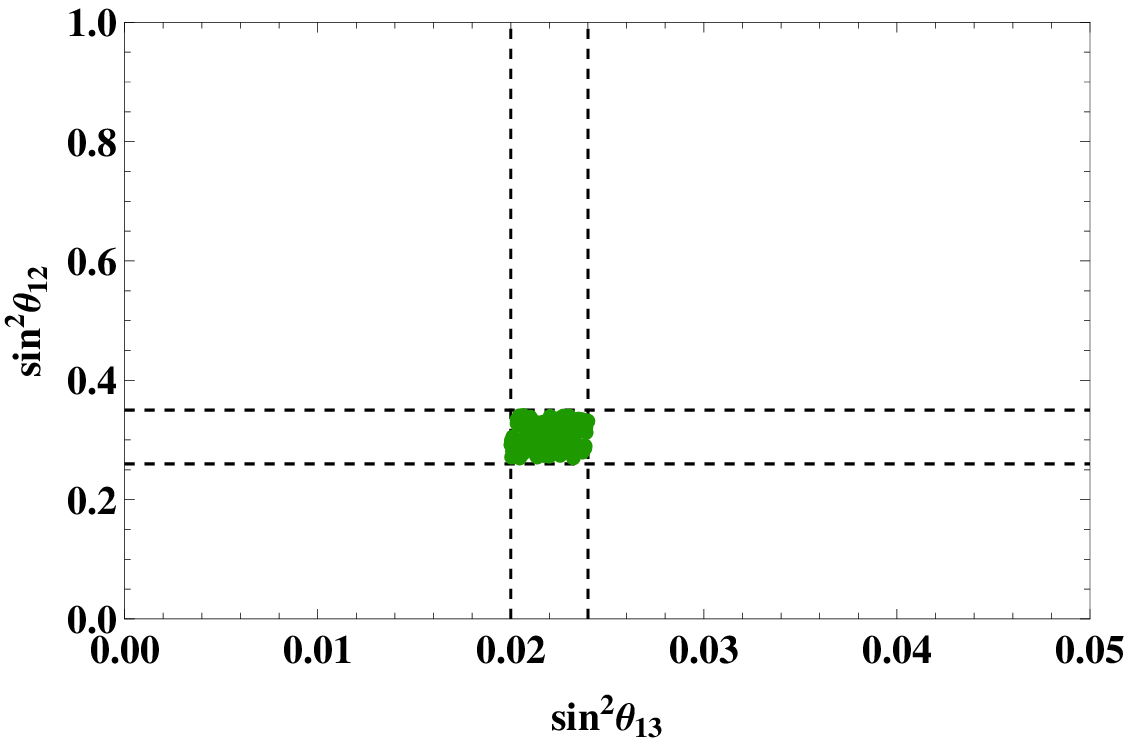}
 \caption{}
 \end{subfigure}
 ~\qquad
 \hspace{.5cm}
 \begin{subfigure}[b]{0.4\textwidth}
 \includegraphics[scale=.7]{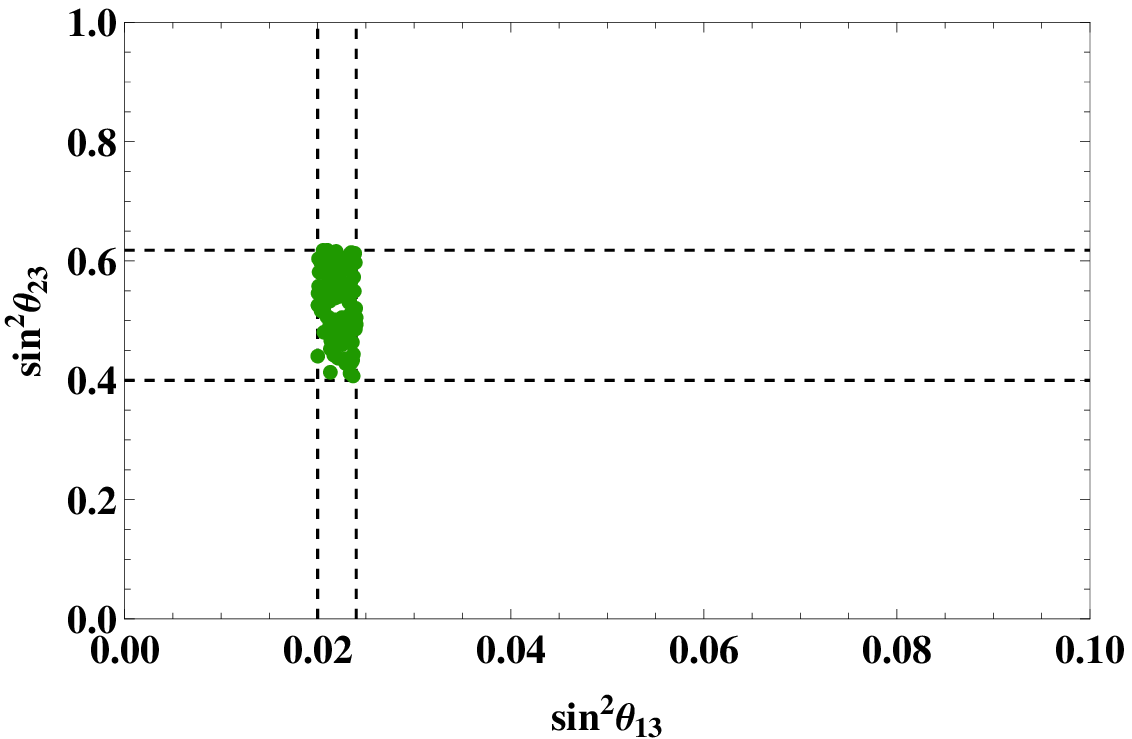}
 \caption{}
 \end{subfigure}
 
 \vspace{1cm}
 \begin{center}
\hspace{-2.5cm}
 \begin{subfigure}[b]{0.4\textwidth}
 \includegraphics[scale=.7]{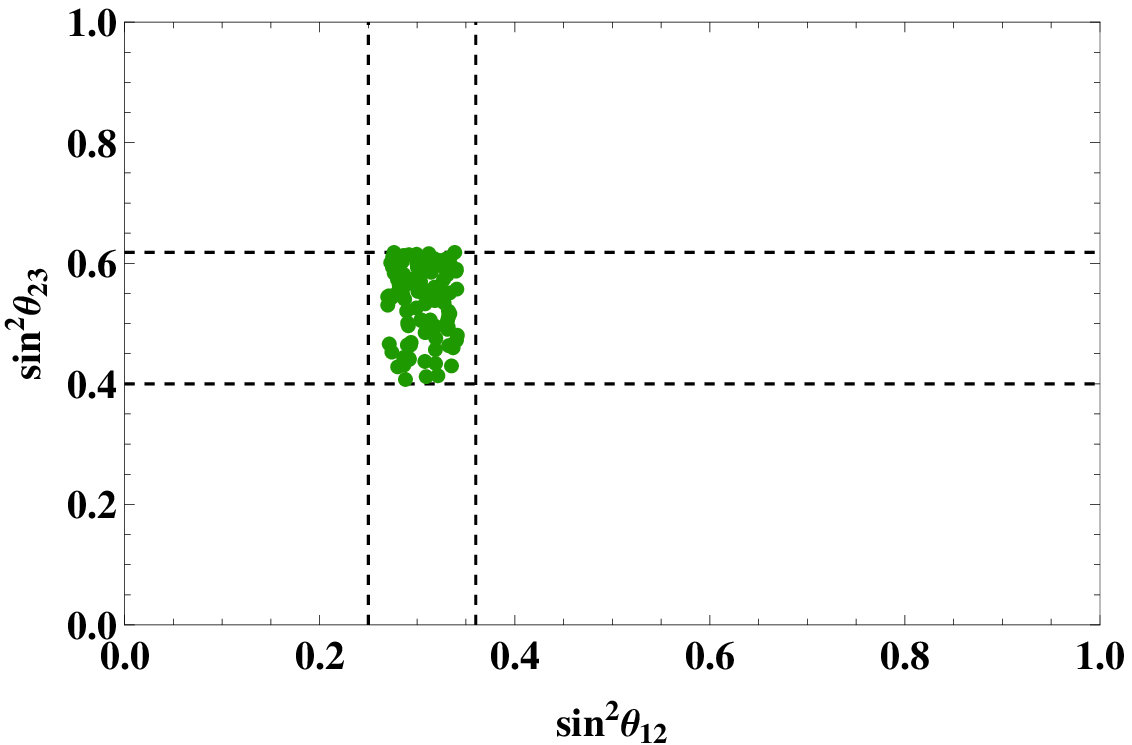}
 \caption{}
 \end{subfigure}
 \end{center}
 \caption{Plots showing correlation of (a)$sin^2 \theta_{12}$ vs $sin^2 \theta_{13}$, (b) $sin^2 \theta_{23}$ vs $sin^2 \theta_{13}$ and (c) $sin^2 \theta_{23}$ vs $sin^2 \theta_{12}$.}
 \label{Fig1}
 \end{figure}

\begin{figure}[ht]
 \begin{subfigure}[b]{0.4\textwidth}
 \includegraphics[scale=.67]{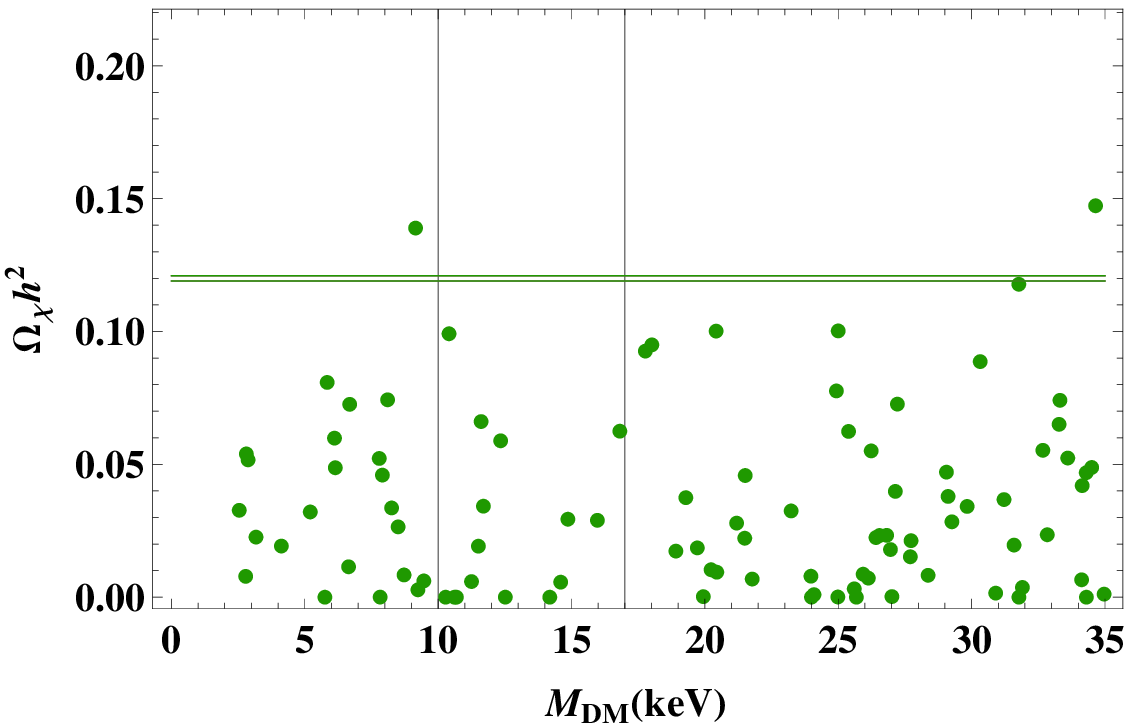}
 \caption{}
 \end{subfigure}
 ~\qquad
 \hspace{.5cm}
 \begin{subfigure}[b]{0.4\textwidth}
 \includegraphics[scale=.67]{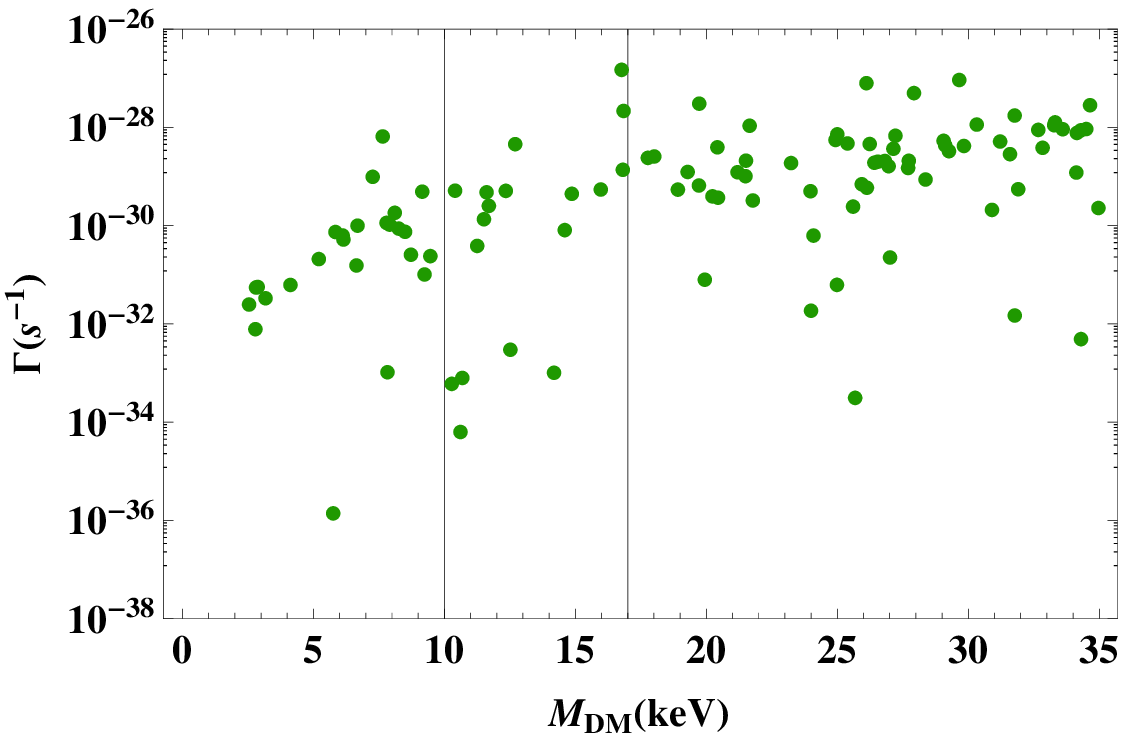}
 \caption{}
 \end{subfigure}
 
 \vspace{1cm}
 \begin{center}
\hspace{-2.5cm}
 \begin{subfigure}[b]{0.4\textwidth}
 \includegraphics[scale=.67]{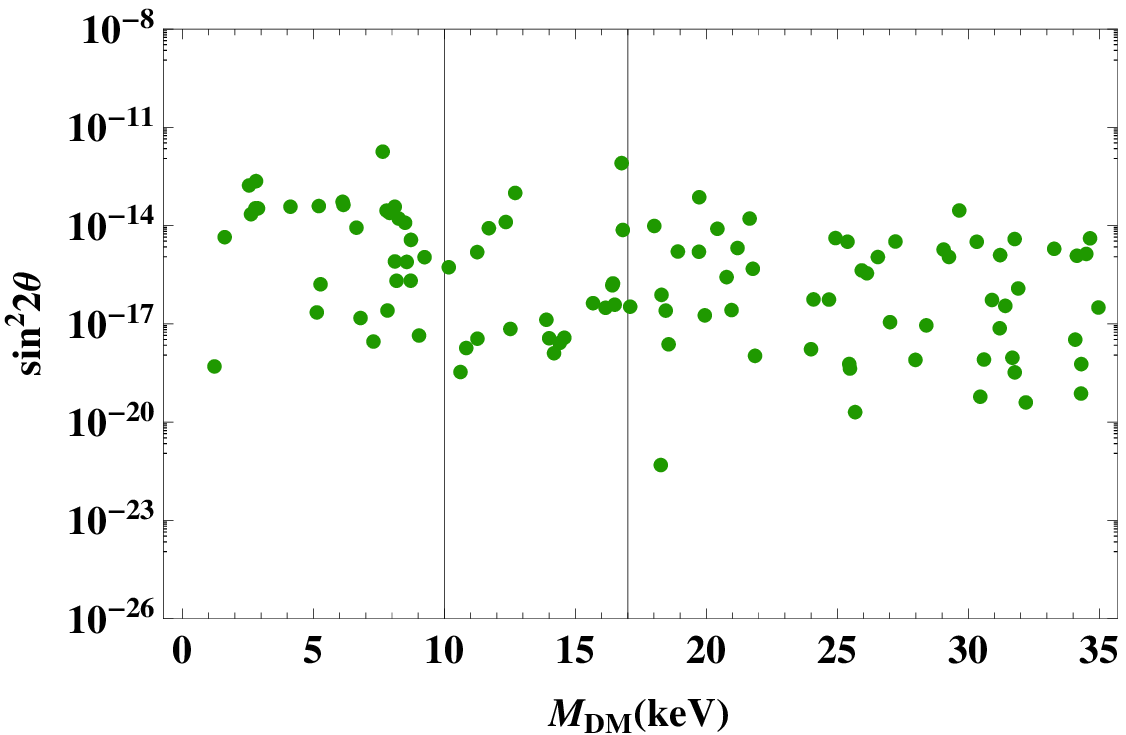}

 \end{subfigure}
 \end{center}
 \caption{Plots showing the variation of (a) Relic density of DM with DM mass , (b) Decay rate of DM with DM mass and (c) active-sterile mixing as a function DM mass.}
 \label{Fig4}
 \end{figure}

\begin{figure}[ht]		\includegraphics[scale=0.7]{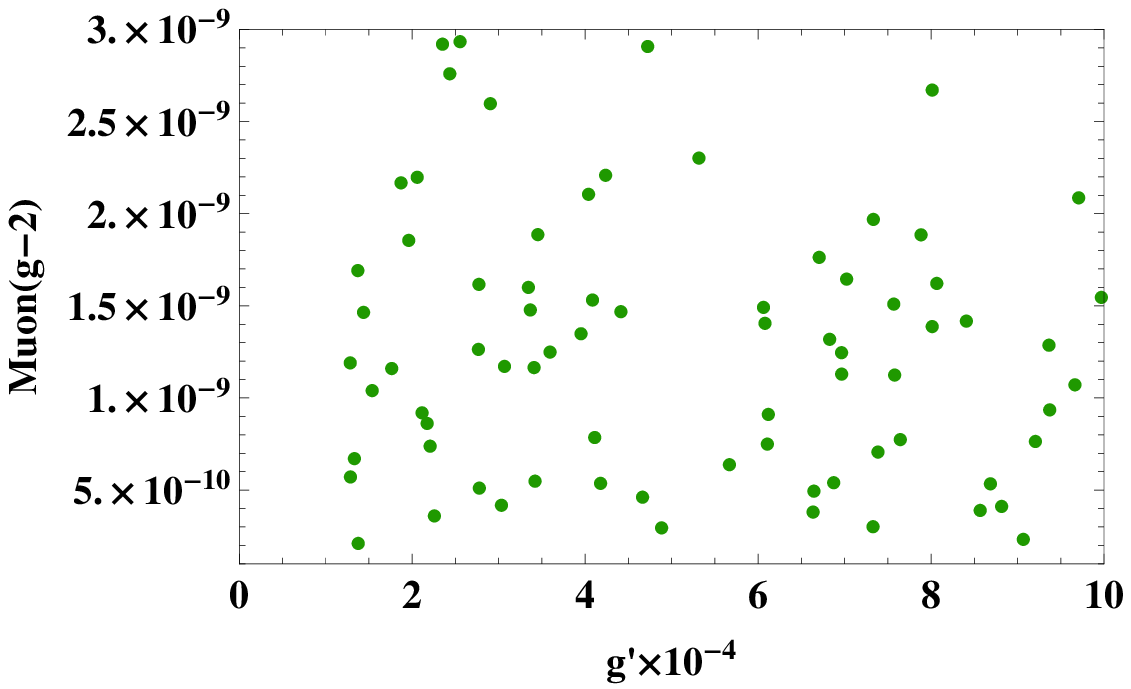}
\includegraphics[scale=0.7]{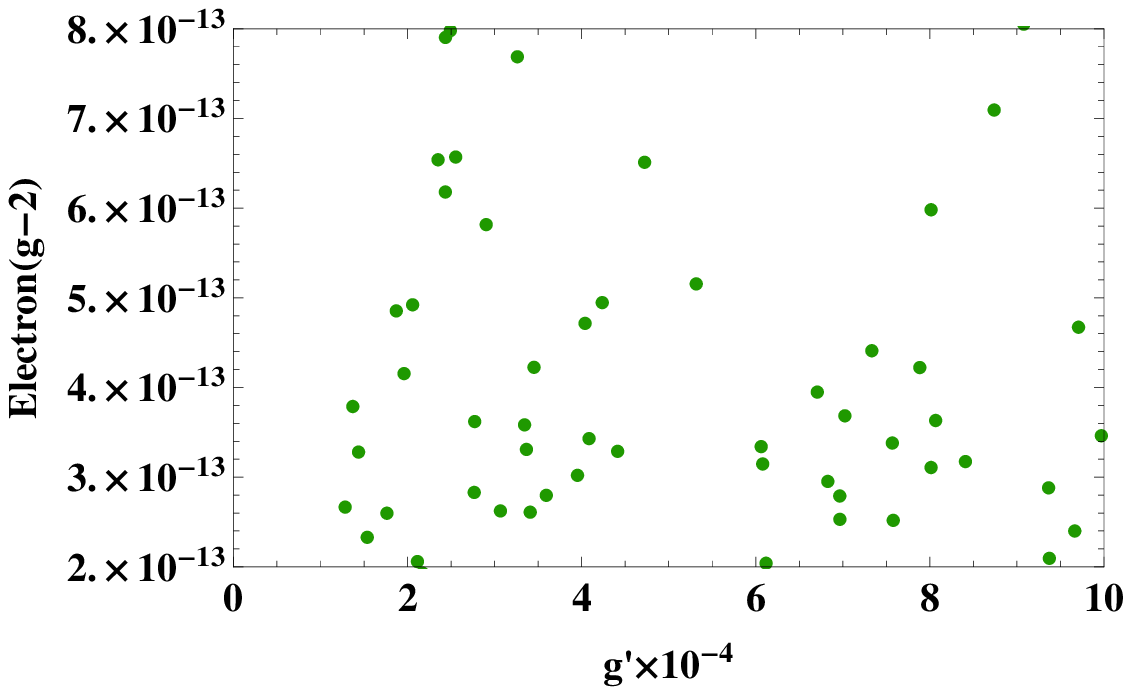}
\caption{\label{fig:4}  Variation of electron and muon anomalous magnetic moment with gauge coupling constant $g'$.}
\label{Fig3}
\end{figure}
\newpage
\noindent The Fig. \ref{Fig1} shows the correlation plots of (a) $sin^2 \theta_{12}$ vs $sin^2 \theta_{13}$ , (b) $sin^2 \theta_{23}$ vs $sin^2 \theta_{13}$ and (c) $sin^2 \theta_{23}$ vs $sin^2\theta_{12}$, for normal ordering. It is evident from Fig. \ref{Fig1}, that the model satisfy the correct neutrino oscillation data on neutrino masses and mixing.

\noindent The Fig. \ref{Fig4} shows the variation of relic density of DM candidate(lightest sterile neutrino), decay rate of DM candidate and active-sterile mixing as a function DM candidate mass $M_{DM}$. Fig. \ref{Fig4}(a) shows the relic abundance of DM candidate as a function of DM mass. The relic abundance obtained in our model satisfy the experimental range for the DM mass within the range (3-35) keV for normal hierarchy. According to the cosmological limits given by Layman-$\alpha$ and X-ray measurements, the DM mass below 10 keV and above 17 keV is excluded, respectively. The the relic abundance shows the partial contribution to the total relic abundance of DM, provided we incorporate these cosmological limits as shown as vertical lines in Fig. \ref{Fig4}, then the relic abundance obtained . Fig. \ref{Fig4}(b) gives the parameter space of decay rate of DM candidate and DM mass. It is clear that the decay rate obtained is negligible and is within the range ($10^{-27}-10^{-36}$) s$^{-1}$ for DM mass range of (2-35) keV. Also, we have shown the active-sterile mixing as a function of DM mass in Fig. \ref{Fig4}(c). In order to be a good DM candidate the sterile neutrino mass must be within the range (0.4-50) keV and its mixing with active neutrinos must be very small and within the range ($10^{-12}-10^{-8}$). It can be seen in the Fig. \ref{Fig4}(c), that the mass and mixing(active-sterile) obtained in our model lies in these ranges.

\noindent In Fig. \ref{Fig3}, we have shown the variation of anomalous magnetic moment with mass of gauge coupling $g'$. The left penal gives the variation of muon anomalous magnetic moment $\Delta a_{\mu}$ with gauge coupling $g'$. The right penal gives the variation of electron anomalous magnetic moment with gauge coupling $g'$. The contribution to anomalous magnetic moment of muon and electron solves the observed discrepancy in contrast to SM.

\section{Conclusions}\label{sec:7}
In this work, we have presented a detailed study of neutrino phenomenology, dark matter, electron and muon (g-2) in an extended SM scenario incorporating ISS(2,3) seesaw mechanism.
Here, we have employed ISS(2,3) seesaw mechanism because it results in an additional keV range sterile state which act as a viable DM candidate. Also, the mass of RHNs are obtained in TeV scale which are accessible at LHC. The model includes two extra RHNs $N_i(i =1,2)$ and three singlet fermions $S_i(i =1,2,3)$ as required in ISS(2,3). The anomaly free $U(1)_{L_{e}-L{\mu}}$ gauge symmetry is used to get an additional $Z'$ gauge boson in MeV range, so that the model can explain the anomalous magnetic moment of electron and muon$(g-2)_{e,\mu}$, simultaneously. Assuming, the lightest sterile neutrino as our DM candidate, we successfully obtained the relic abundance of DM. The calculated relic abundance satisfies the experimental range for the DM mass range (3-35)keV. Further, we have calculated the decay rate of DM candidate to check its stability. We found that the DM candidate is stable as we have obtained negligible decay rate. Also, the active-sterile mixing and DM mass are compatible with the experimental ranges. 

\noindent In summary, we developed a model formalism which could explain the correct neutrino phenomenology, DM problem and anomalous magnetic moment of electron and muon$(g-2)_{e,\mu}$.

\hspace{-.4cm}\textbf{\Large{Acknowledgments}}
 \vspace{.3cm}\\
 R. Verma acknowledges the financial support provided by the Central University of Himachal Pradesh. B. C. Chauhan is thankful to the Inter University Centre for Astronomy and Astrophysics (IUCAA) for providing necessary facilities during the completion of this work. Ankush acknowledges the financial support provided by the University Grants Commission, Government of India vide registration number 201819-NFO-2018-19-OBCHIM-75542. Special thanks to Monal Kashav for his invaluable input and constant support throughout the research process.

\end{document}